\documentclass[prb, preprint, amsmath, amssymb, showpacs, superscriptaddress,longbibliography]{revtex4-1}
\usepackage{graphicx} 
\usepackage{dcolumn}  
\usepackage{bm}       
\usepackage{amsfonts}
\usepackage{dsfont}
\usepackage{csquotes}

\usepackage{ulem}
\usepackage{color}

\usepackage{float}
\usepackage{physics}
\usepackage{comment}

\usepackage{hyperref}
\begin{document}

\title{Tailoring Photocurrent in Weyl Semimetals via Intense Laser Irradiation}

\author{Amar Bharti}
\affiliation{%
	Department of Physics, Indian Institute of Technology Bombay,
    Powai, Mumbai 400076, India }

\author{Gopal Dixit}
\email[]{gdixit@phy.iitb.ac.in}
\affiliation{%
	Department of Physics, Indian Institute of Technology Bombay,
	Powai, Mumbai 400076, India }
\affiliation{%
Center for Computational Sciences, University of Tsukuba, Tsukuba 305-8577, Japan}
\affiliation{%
Max-Born Institut, Max-Born Stra{\ss}e 2A, 12489 Berlin, Germany}

\date{\today}


\begin{abstract}
Generating and tailoring photocurrent  in topological materials has immense importance 
in fundamental studies and the technological front. 
Present work introduces a universal method to generate ultrafast photocurrent in
{\it both} inversion-symmetric and inversion-broken Weyl semimetals with 
degenerate Weyl nodes at the Fermi level.
Our approach harnesses the asymmetric electronic population in the conduction band induced by an intense  
{\it single-color} circularly polarized laser pulse.  
It has been found that the induced photocurrent can be tailored 
by manipulating helicity and ellipticity of the employed laser.  
Moreover, our approach generates photocurrent in realistic situations  
when the Weyl nodes are positioned at different energies and have finite tilt along a certain direction.
Present work adds a new dimension on  practical applications of  
Weyl semimetals for optoelectronics and photonics-based quantum technologies.
\end{abstract}

\maketitle

Weyl semimetals  are topological materials that have demonstrated the potential 
to convert light into electricity efficiently. 
The superiority of the Weyl semimetal over many materials in generating photocurrent in 
the infrared region has been established experimentally~\cite{osterhoudt2019colossal}. 
In addition,  ultrafast photocurrent from  Weyl semimetals can be a source of terahertz radiation~\cite{sirica2019tracking,gao2020chiral}.  
Moreover,   photocurrent emerges as a  quintessential probe of the topological properties of quantum materials
~\cite{pan2017helicity, zhang2019switchable, wang2020electrically}, including device characterization~\cite{ma2023photocurrent}.  
Thus, recent developments in producing  photocurrent from Weyl semimetals make them  
a central focus for various applications in optoelectronics,  
detection, and sensing to name but a few~\cite{wang2019robust,luo2021light,ma2022unveiling,wang2023visualization,bharti2023role, liu2020semimetals, ma2019nonlinear}. 

Asymmetric population distribution of the  electronic excitations in Weyl semimetal renders a finite photocurrent -- photogalvanic effect -- 
which can be realized in various ways, such as the chiral magnetic effect~\cite{taguchi2016photovoltaic,kaushik2019chiral,kaushik2020transverse}, 
 via transfer of angular momentum of light to the Weyl nodes~\cite{de2017quantized}, 
and nonlinear optical responses in the perturbative regime~\cite{wang2020electrically,watanabe2021chiral,heidari2022nonlinear,golub2017photocurrent,sirica2022photocurrent,golub2018circular,golub2017photocurrent,ishizuka2016emergent, zhang2018photogalvanic, gao2021intrinsic}.  
It has been shown that the inversion-broken  Weyl semimetal with gyrotropic symmetry produces 
second-order nonlinear optical responses -- 
injection and shift currents -- which lead to a colossal photocurrent in TaAs ~\cite{morimoto2016topological,orenstein2021topology,ma2021topology, osterhoudt2019colossal}. 
In addition, photocurrent can exhibit a sign flip with the change in the helicity of the circularly polarized light~\cite{de2017quantized,chan2017photocurrents,konig2017photogalvanic}. 
The broken mirror-symmetry of Weyl nodes  
is a key reason 
behind helicity-sensitive photoresponse in the inversion-broken Weyl semimetals
~\cite{ma2017direct,rees2020helicity,ni2021giant}. 
So far, the majority of the work on photocurrent is focused on inversion-broken Weyl semimetals with various tilts and crystal symmetries, as in the recent one in an inversion-symmetric  Weyl semimetal ~\cite{hamara2023helicity}. 
Thus, a universal method  to generate photocurrent from   both inversion-symmetric and inversion-broken   Weyl semimetals that does not rely on such materials' symmetry details is lacking.

It is a commonly accepted notion that the single-color circularly polarized light fails to generate 
photocurrent in 
Weyl semimetals with mirror-symmetric Weyl nodes. 
While each Weyl node generates current depending on its chirality, 
the currents in a chiral pair of the mirror-symmetric Weyl nodes cancel each other. 
In contrast to this accepted notion, we unequivocally demonstrate that a single-color circularly 
polarized light is able to generate photocurrent in mirror-symmetric Weyl semimetals.
Our approach does not rely on the system's symmetry 
as it is  equally applicable to 
both inversion-symmetric and inversion-broken Weyl semimetals with isotropic band dispersion 
and even with all Weyl nodes  at the Fermi level.
Recently, bicircular laser pulses have been proposed for generating  photocurrent in two- and three-dimensional materials, including  Weyl semimetal, described by a linear anisotropic Hamiltonian~\cite{ikeda2023photocurrent, neufeld2021light}. 
However, a single-color laser based photocurrent, with 
relatively easy experimental setup, 
is highly desirable for practical purposes. 
Few-cycle carrier-envelope phase stabilized linearly polarized laser 
can induce photocurrent as shown for graphene~\cite{higuchi2017light,zhang2022bidirectional}. 
However, such photocurrent cancels out if the phase is not stabilized, outweighing its applicability.
Our approach is also robust against such carrier-envelope phase stabilization.
We employ three- and six-cycle laser pulses in the mid-infrared regime  
to generate photocurrents whose direction and magnitude can be tailored by the phase of the circular pulse.
Moreover, it is observed that the photocurrent in an inversion-symmetric Weyl semimetal is sensitive to the helicity and ellipticity of the laser pulse.

We start our discussion by writing the Hamiltonian of Weyl semimetals as 
$\mathcal{H}(\mathbf{k}) = \mathbf{d}(\mathbf{k}) \cdot \sigma$, 
with $\sigma$'s being the Pauli matrices. 
Expressions of  the three components of $\mathbf{d}(\mathbf{k})$
for an inversion-symmetric Weyl semimetal are~\cite{sadhukhan2021role,menon2021chiral} 
\begin{equation}\label{eq:trb}
\mathbf{d}(\mathbf{k})  =   \big[t\sin(k_x a), t\sin(k_y a),
 t\{\cos(k_z a) - \cos(k_0 a) +2- \cos(k_x a) - \cos(k_y a)\}\big],
\end{equation}
and for an inversion-broken Weyl semimetal read as 	 
\begin{eqnarray}\label{eq:invb}
&\mathbf{d}(\mathbf{k})  =  & \big[t\{\cos(k_0 a)-\cos(k_y a) + \mu[1-\cos(k_z a)]\}, t\sin(k_z a),\nonumber \\
 && t\{\cos(k_0 a)-\cos(k_x a) + \mu[1-\cos(k_z a)]\}\big].
\end{eqnarray}
Here, $k_0$ determines the position of the Weyl nodes, which are considered as $\pi/(2a)$ for both Weyl semimetals. 
The Weyl nodes for inversion-symmetric and inversion-broken systems are situated at $\mathbf{k} = [0,0,\pm \pi/(2a)]$ and $\mathbf{k}=[\pm\pi/(2a),\pm\pi/(2a),0]$, respectively.  
 A simple cubic crystal structure is considered with lattice parameter $a = 6.28$ \AA ~and isotropic hopping parameter $t=1.8$ eV  
in Eqs.~(\ref{eq:trb}) and~(\ref{eq:invb}). 
Moreover, a dimensionless parameter $\mu = 2$ is used in Eq.~(\ref{eq:invb}). 
Energy band dispersions corresponding to Eqs.~(\ref{eq:trb}) and  (\ref{eq:invb}) are shown in Figs.~S1 and S2 in Ref.~\cite{NoteX}, respectively.

The vector potential of the circularly polarized laser is written as 
$\mathbf{A}(t) = A_0 \Re{e^{i \omega t + \phi} ~\hat{\mathbf{e}}_{\pm}}$,  
where $\hat{\mathbf{e}}_{\pm} = (\hat{\mathbf{e}}_{x} \pm i \epsilon\ \hat{\mathbf{e}}_{y})$ corresponds to left- and right-handed circularly polarized laser pulse with ellipticity $\epsilon=1$. 
The subcycle phase of the laser pulse is denoted by $\phi$, which controls the orientation of the Lissajous profile of the laser. 
A laser pulse having a sine-squared envelope with  wavelength $3.2 ~\mu$m  
and pulse duration ranging from $\sim$ 35 to 70 fs is employed to generate photocurrent. 
The density-matrix-based approach is used to simulate laser-driven dynamics in Weyl semimetals as discussed in Refs.~\cite{mrudul2021high,wilhelm2021semiconductor, rana2022high, bharti2022high}.
The  photocurrent originates from the population asymmetry and can be written as
~\cite{soifer2019band}
\begin{equation}\label{eq:1}
\mathbf{J}(t) = \int_\mathbf{k}  d\mathbf{k}~\left[ \rho(\mathbf{k}) - \rho(-\mathbf{k}) \right] \pdv{\mathcal{E}(\mathbf{k})}{\mathbf{k}},
\end{equation}
where $\mathbf{J}(t)$ is the total current, $\rho$ is the residual  population  density  
after the end of the laser pulse, 
and  $\mathcal{E}(\mathbf{k})$ is the energy dispersion in a Weyl semimetal.

\begin{figure}[!htb]
\centering
\includegraphics[width=\linewidth]{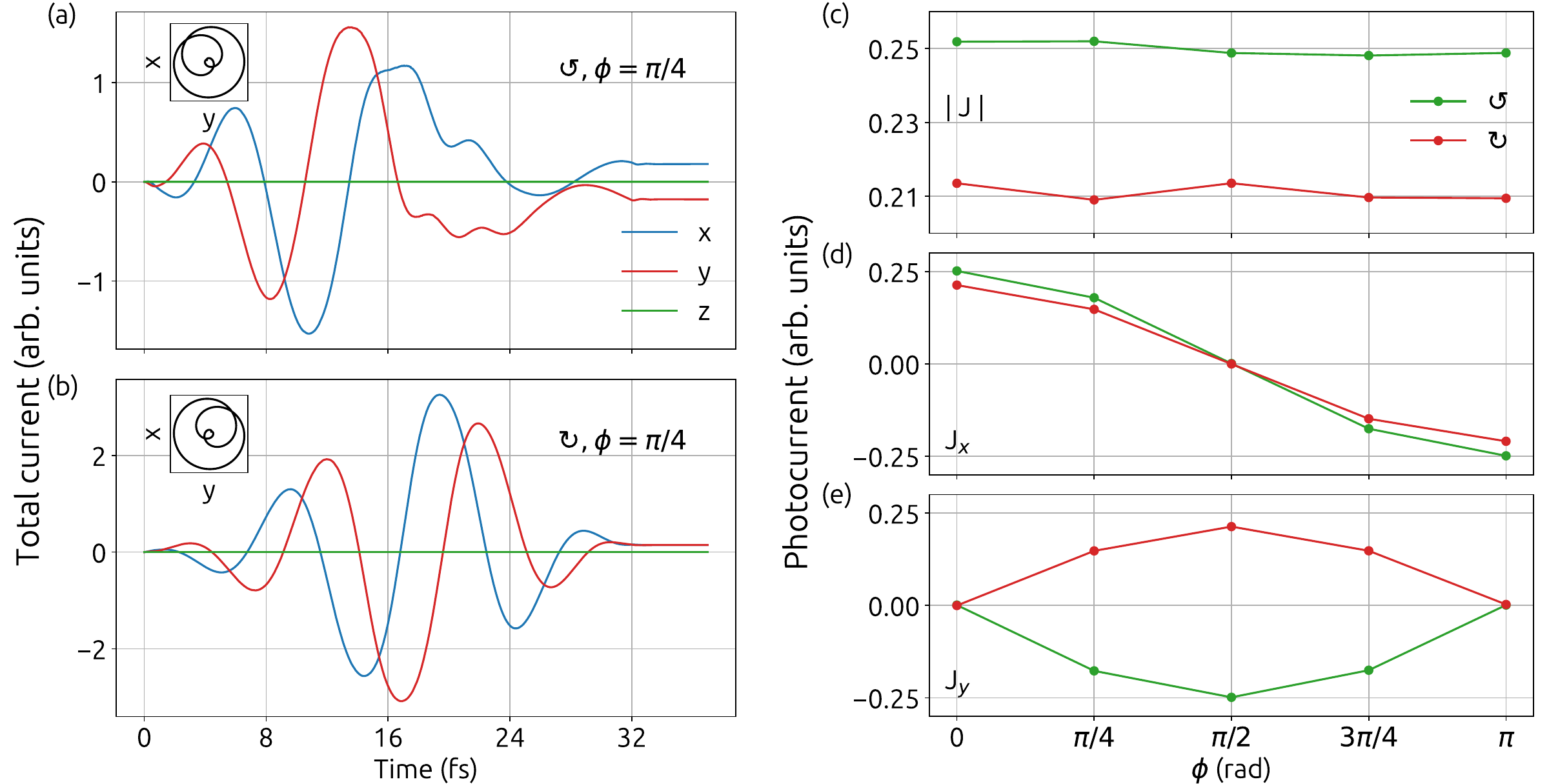}
\caption{Total photocurrent in an inversion-symmetric Weyl semimetal 
induced by (a)  left-handed and (b) right-handed circularly polarized laser with  phase 
$\phi = \pi/4$. The Lissajous curve in the $xy$ plane is shown in respective insets. 
Variations in (c) the   photocurrent, (d) its $x$  ($\mathsf{J}_{x}$), and (e) $y$  
($\mathsf{J}_{y}$) components with respect to the phase. 
These results are obtained for a three-cycle circularly polarized laser pulse with  $\sim$ 32 fs duration, 3.2 $\mu m$ wavelength, and $10^{11}$ W/cm$^2$ intensity.}
\label{fig:fig1}
\end{figure}

Let us analyze results for an inversion-symmetric Weyl semimetal, which exhibits 
a finite photocurrent  along $x$ and $y$ directions after the end of the laser pulse 
as shown in  Fig.~\ref{fig:fig1}(a). 
As the helicity of the laser changes from right to left, the sign of the photocurrent  along the $y$ direction  flips  
from negative to positive as evident from Fig.~\ref{fig:fig1}(b). 
To unravel the underlying mechanism  for the flip, we analyzed the Lissajous profile of the vector potential 
in the polarization plane as shown in the insets. 
The change in the Lissajous curve with a change in the helicity is a primary 
 reason for the sign flips along  $y$ direction. 
This shows that the photocurrent is susceptible to the profile of the laser pulse. 
At this juncture, it is pertinent to know how the photocurrent is sensitive to the phase of the laser pulse. 
To this end, we investigate variation in the total photocurrent and its components with respect to  the phase.

\begin{figure}[!htb]
\centering
\includegraphics[width=\linewidth]{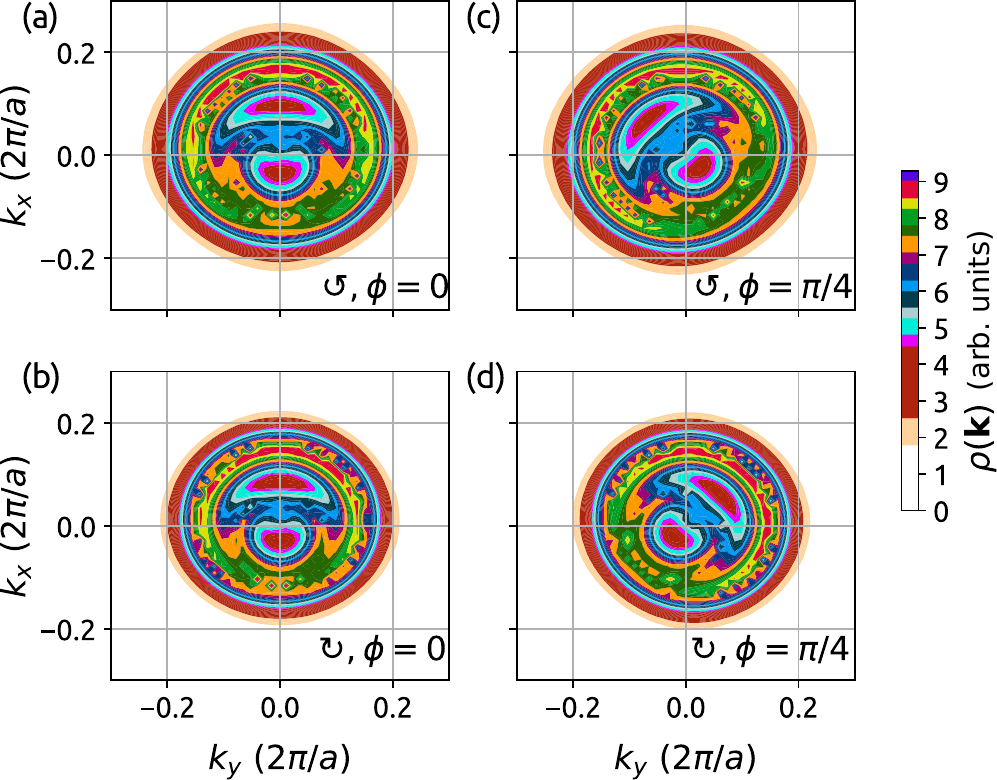}
\caption{Residual population in the conduction band  of an inversion-symmetric Weyl semimetal 
after the end of the  (a) left-handed, and (b) right-handed circularly polarized laser with $\phi =0$. 
(c) and (d) are, respectively the same as (a) and (b) for $\phi = \pi/4$. 
Weyl nodes are situated at 
$\mathbf{k}=[0,0, \pm \pi/(2a)]$ according to Eq. (\ref{eq:trb}). 
Population along $k_z$ direction is integrated in all cases. Laser parameters are identical to  Fig. \ref{fig:fig1}. }
\label{fig:fig2}
\end{figure}

Figure~\ref{fig:fig1}(c) shows the insensitivity of the photocurrent with respect to $\phi$, 
which concludes that phase stabilization is not a prerequisite to generate photocurrent in Weyl semimetal.  
However,  the $x$ component ($\mathsf{J}_{x}$) changes from a  positive  to 
a negative value as $\phi$ changes from 0 to $\pi$, including zero at $\phi = \pi/2$ [see Fig.~\ref{fig:fig1}(d)]. 
Both helicities display similar behavior for $\mathsf{J}_{x}$, whereas 
the $y$ component ($\mathsf{J}_{y}$) exhibits an opposite trend as the helicity is reversed 
from left and right, except $\phi = 0$ and  $\pi$ where it is zero [see Fig.~\ref{fig:fig1}(e)].  
Analysis of Fig.~\ref{fig:fig1} raises a crucial question 
about factors determining the nonzero photocurrent and its components.  

The residual population in the conduction band around a Weyl node after the end of the laser is presented in Fig.~\ref{fig:fig2}. 
Owing to the zero band-gap nature of the Weyl node, the region around the node is significantly populated, which decreases rapidly as we move away from the origin. 
Population about  the $k_{x} = 0$ plane is significantly asymmetric, which in results nonzero 
photocurrent along this direction as 
$\rho({k_{x}}) \neq \rho(-{k_{x}})$ for both helicities. 
However, the population exhibits mirror symmetry about the $k_{y} = 0$ plane for $\phi = 0$, which in results in zero 
photocurrent for both helicities as evident from Figs.~\ref{fig:fig1}(e),  
\ref{fig:fig2}(a) and \ref{fig:fig2}(b). 
A change in  $\phi$  from 0 to $\pi/4$ induces asymmetry along $k_{y} = 0$, which generates finite photocurrent as 
reflected from Figs.~\ref{fig:fig2}(a) and \ref{fig:fig2}(b). 
In addition, the direction of the induced asymmetry along $k_{y} = 0$ flips as we change the helicity from left and right, which results in a sign change in  $\mathsf{J}_{y}$ as shown in Fig.~\ref{fig:fig1}(e). 
Thus, observations 
in Figs.~\ref{fig:fig1}  and \ref{fig:fig2} are consistent  with Eq.~(\ref{eq:1}). 

\begin{figure}[!htb]
\centering
\includegraphics[width=\linewidth]{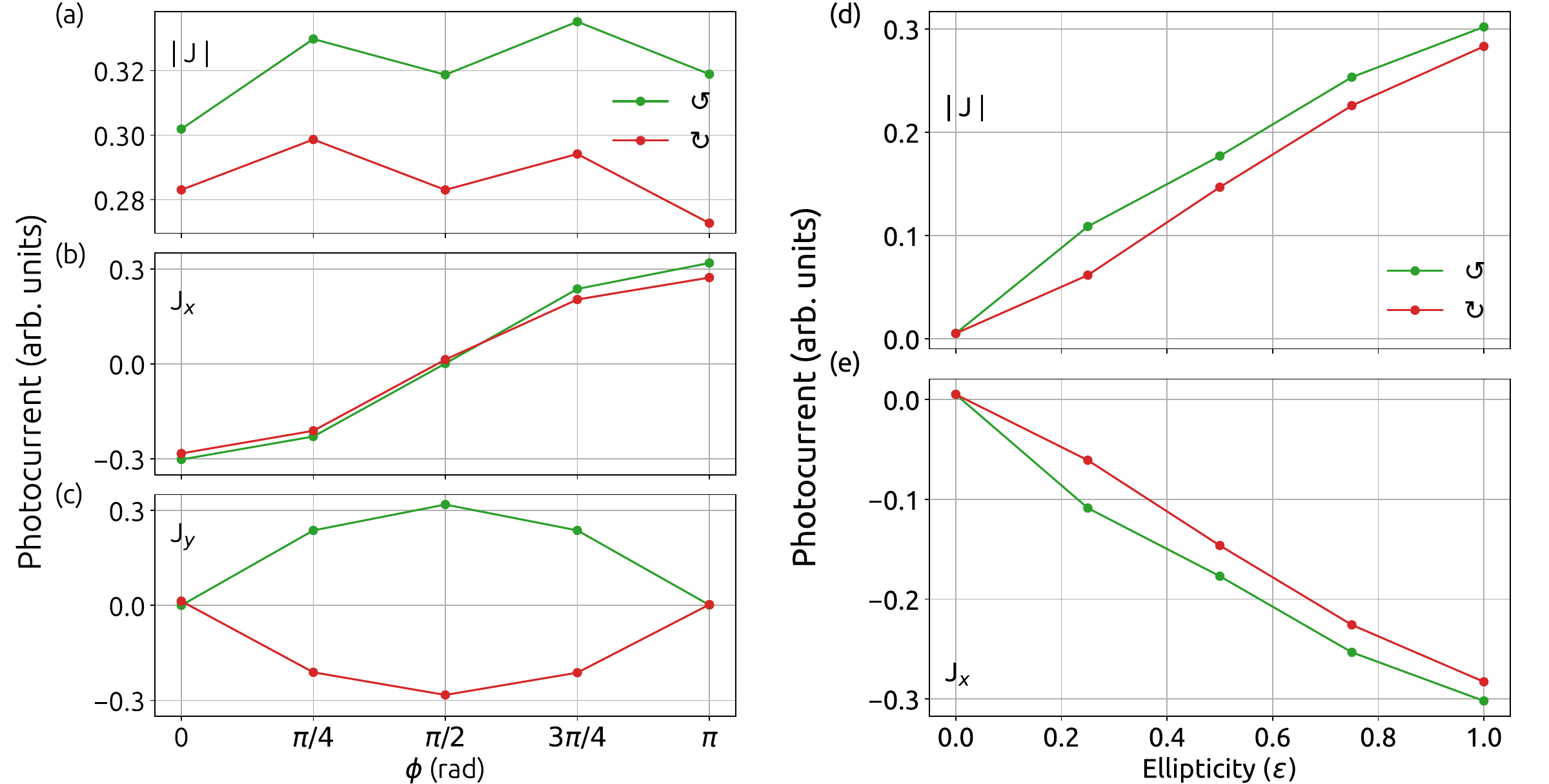}
\caption{Variations in (a) the  total photocurrent,  
(b) its $x$  ($\mathsf{J}_{x}$), and (c) $y$  
($\mathsf{J}_{y}$) components with respect to $\phi$ for an inversion-symmetric Weyl semimetal.  
Sensitivity of (d) the photocurrent and (e)   $\mathsf{J}_{x}$  with respect to the ellipticity of the laser for
$\phi=0$.  The laser pulse consists of six cycles with  $5\times10^{11}$ W/cm$^2$ intensity and 3.2 $\mu m$ wavelength.}
\label{fig:fig3}
\end{figure}

One of the striking features of Fig.~\ref{fig:fig2} is the extent of the asymmetries along $k_{x} = 0$ and  $k_{y} = 0$ planes, which are significantly different for both helicities. 
Recently, it has been shown that the electronic excitation from the nonlinear part of the band dispersion 
can effectuate the helicity-dependent population in an inversion-symmetric Weyl semimetal~\cite{bharti2023weyl}. 
Therefore, owing to the unique coupling of the circularly polarized laser with the Weyl semimetal,  
the residual population along $k_z$, integrated along other directions, is sensitive to the laser's helicity~\cite{bharti2023weyl}. 
Thus, the helicity-sensitive population asymmetry leads to different photocurrent for the left- and right-handed laser pulses as shown in Fig.~\ref{fig:fig1}.

So far, we have discussed the results of the three-cycle laser pulse. 
It is known that the vector potential can be nonzero when the electric field is  zero for a  
few-cycle laser pulse with stabilized carrier-envelope phase. 
The nonzero vector potential can induce asymmetric population and photocurrent in graphene, as discussed in Refs.~\cite{higuchi2017light,zhang2022bidirectional}. 
Moreover, the resultant asymmetric population can also yield valley polarization in two-dimensional materials~\cite{rana2023all, mrudul2021light, mrudul2021controlling}.  
Thus, it is natural to ask about the robustness of our results with the  pulse duration.
Generating photocurrent in Weyl semimetals via relatively long laser pulse in mid-infrared regime is highly desirable for numerous practical  applications~\cite{wang2020electrically,watanabe2021chiral,heidari2022nonlinear,golub2017photocurrent,sirica2022photocurrent,golub2018circular,golub2017photocurrent,ishizuka2016emergent}. 

Towards that end, let us  increase the pulse duration from $\simeq$   30 to 65 fs by changing 
the number of cycles from three to six while keeping the intensity constant.   
In this case, a finite photocurrent with a relatively smaller magnitude is observed.  
It is found that the intensity needs to be increased by five times 
to make the magnitude of the photocurrent comparable for three- and six-cycle pulses [see Fig.~\ref{fig:fig3}(a)]. 
On comparing Figs.~\ref{fig:fig1}(c) and \ref{fig:fig3}(a), it is evident 
that an increase in intensity leads to a reduction in the contrast between the photocurrent for different helicity. 
The reduction in the contrast can be attributed to the underlying mechanism of the helicity-dependent asymmetric population, which relies on the resonant excitation at various $\mathbf{k}$ and thus reduces the asymmetry with an increase in intensity~\cite{bharti2023weyl}. 

In contrast to the three-cycle pulse, $\mathsf{J}_{x}$ transits from negative to positive magnitude as $\phi$ changes from 
0 to $\pi$, whereas $\mathsf{J}_{y}$ exhibits similar behavior for three- and six-cycle pulses [see Figs.~\ref{fig:fig3}(b) and \ref{fig:fig3}(c)]. 
Note that the photocurrent 
can be positive or negative based on whether $-\mathbf{k}$ or $\mathbf{k}$ is more populated, which 
depends on the intensity and pulse duration~\cite{zhang2022bidirectional}. 
The photocurrent is not only sensitive to the pulse duration but also to the ellipticity of the laser pulse, as shown in Fig.~\ref{fig:fig3}(d) for $\phi =0$. 
Photocurrent monotonically reduces to zero as the ellipticity changes from one (circular)  to zero (linear) for both helicities. Similar observations can be made for $\mathsf{J}_{x}$ from Fig.~\ref{fig:fig3}(e). 
Note that $\mathsf{J}_{y}$ is zero for $\phi =0$. 
The generated photocurrent is nonperturbative in nature as evident from its scaling with laser's intensity  [see Fig. S3~\cite{NoteX}].
Our analysis establishes that a laser pulse with definite chirality, but nonzero ellipticity, 
is able to engender photocurrent in 
an inversion-symmetric Weyl semimetal, which also encapsulates a unique coupling of chiral light with Weyl semimetal~\cite{bharti2023weyl}. 
Our approach is equally applicable to realistic situations when the Weyl nodes are nondegenerate [see Fig. S4~\cite{NoteX}], situated at different energies, and  have tilt along certain direction [see Fig. S5~\cite{NoteX}].
In addition, our method produces photocurrent 
of the same order [see Fig. S6~\cite{NoteX}] 
as the one reported by Morimoto  and coworkers using 
bicircular counter-rotating laser pulses~\cite{ikeda2023photocurrent}.

\begin{figure}[!htb]
\centering
\includegraphics[width=\linewidth]{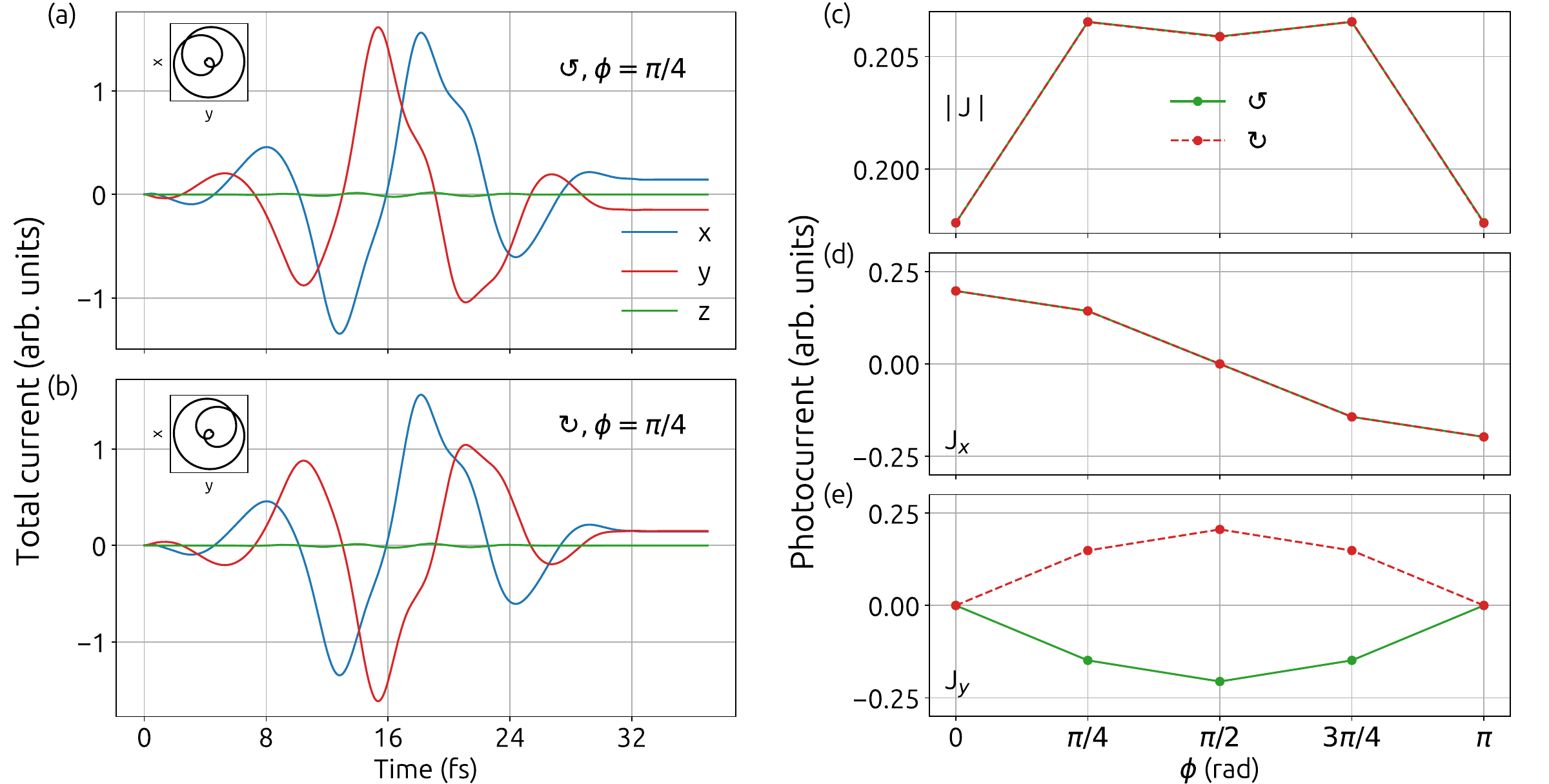}
\caption{Same as Fig. \ref{fig:fig1} for an inversion-broken Weyl semimetal as described  
by the Hamiltonian in Eq. \eqref{eq:invb}.}
\label{fig:fig4}
\end{figure}

After demonstrating the photocurrent generation in an inversion-symmetric Weyl semimetal, 
let us focus our discussion to inversion-broken Weyl semimetal. 
Figure~\ref{fig:fig4} presents finite photocurrent in an inversion-broken Weyl semimetal driven by a circularly polarized laser. 
 By the virtue of the Lissajous profile flip, the photocurrent along $y$ direction flips its sign as the laser's helicity changes for $\phi = \pi/4$ [see Figs.~\ref{fig:fig4}(a) and \ref{fig:fig4}(b)]. 
The total photocurrent does not change significantly with variation in $\phi$ and is identical for both helicities as shown in  Fig.~\ref{fig:fig4}(c). 
Similar to an inversion-symmetric case, 
$\mathsf{J}_{x}$ changes its magnitude from positive to negative as 
$\phi$ changes from 0 to $\pi$ [see Fig.~\ref{fig:fig4}(d)], and $\mathsf{J}_{y}$ remains either positive or negative depending on the helicity 
except at $\phi = 0$ and $\pi$ [see Fig.~\ref{fig:fig4}(e)]. 
Thus, the behavior of the photocurrent and its components are robust with respect to $\phi$. 
Note that there is a finite photocurrent in the plane of polarization for other polarization directions of the laser, and can be tailored by changing $\phi$. 

\begin{figure}[!htb]
\centering
\includegraphics[width=\linewidth]{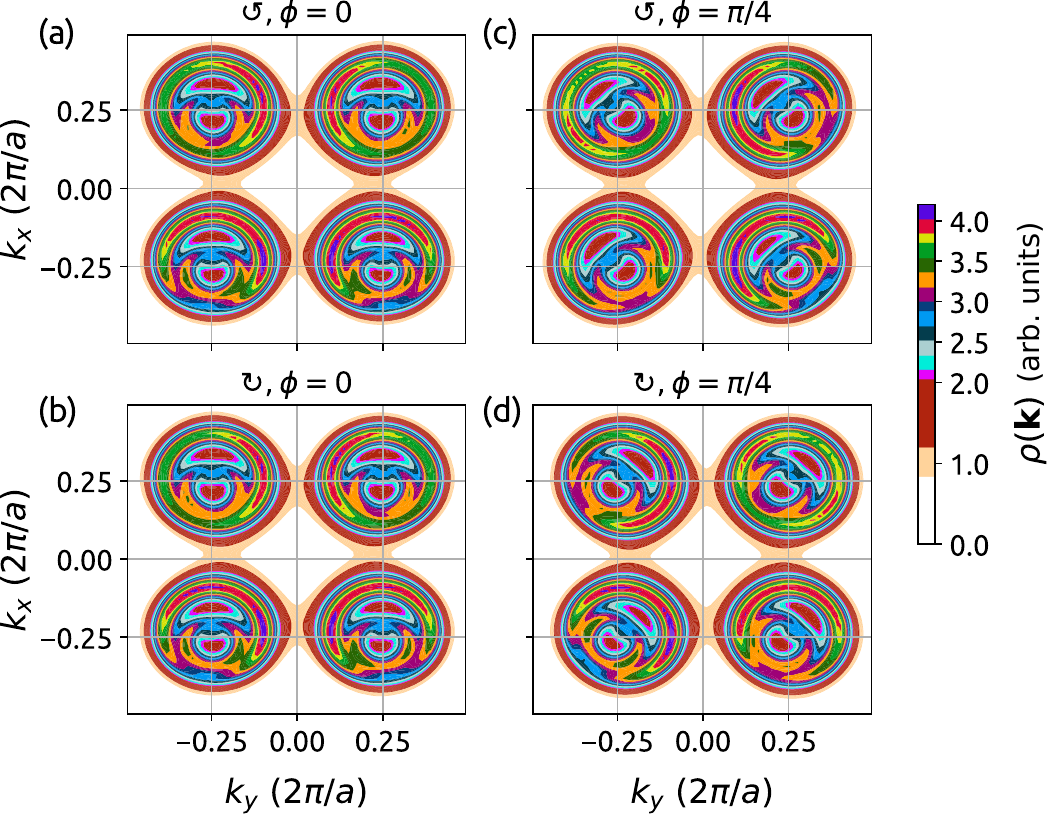}
\caption{Same as Fig. \ref{fig:fig2} for an inversion-broken Weyl semimetal.}
\label{fig:fig5}
\end{figure}

We also analyze  the residual  
population in the conduction band to corroborate  the photocurrent's results  in Fig.~\ref{fig:fig4}. 
Significant  population around four Weyl nodes at $\mathbf{k}=[\pm\pi/(2a),\pm\pi/(2a),0]$ is observed 
as shown in Fig.~\ref{fig:fig5}. 
Population is asymmetric in nature with respect to $k_x=0$ plane for $\phi=0$ 
and exhibits $k_y = 0$ as a plane of reflection, which results in nonzero (zero)  photocurrent along $x$ ($y$) axis.   
Reflection symmetry  about $k_y = 0$  plane is lost as $\phi$ changes  to $\pi/4$ 
[see Figs.~\ref{fig:fig5}(c) and \ref{fig:fig5}(d)],  
which results in nonzero  photocurrent along this direction as evident from Fig.~\ref{fig:fig4}(e). 
In addition, the population corresponding to both helicities are identical, which is in contrast to the one observed for  
an inversion-symmetric Weyl semimetal.  

To summarize, we introduce a robust and universal  method to generate photocurrent in both 
inversion-symmetric and inversion-broken Weyl semimetals using a single-color circularly polarized light.  
Both Weyl semimetals have degenerate Weyl nodes at Fermi level. 
We unequivocally show  that phase stabilization is  not a prerequisite to generate  photocurrent in 
both types of the Weyl semimetals as the generated photocurrent is insensitive to the phase of the laser pulse. 
Photocurrent in an inversion-symmetric  Weyl semimetal is sensitive to the helicity of the laser as the left-handed circularly polarized laser yields more photocurrent in comparison to the right-handed laser. 
Moreover, the components of the photocurrent in an  inversion-symmetric  Weyl semimetal are also 
sensitive to the helicity, whereas 
only the $y$ component exhibits sensitivity in case of an inversion-broken Weyl semimetal. 
In addition, the strength of the photocurrent reduces as the ellipticity of the laser changes  from circular to linear. 
It is anticipated that the measurement of the photocurrent can quantify the coupling of spin-angular momentum of light with nonlinear band dispersion in Weyl semimetals. 
Our introduced method can be extended to other topological materials  for their widespread applications in 
optoelectronics and photonics.

G. D. acknowledges fruitful discussions with  Misha Ivanov (MBI, Berlin) and 
 Kazuhiro Yabana (Tsukuba University). 
G. D. acknowledges financial  support from SERB India  (Project No. MTR/2021/000138).

\newpage
\bibliographystyle{apsrev4-1}
\bibliography{solid_HHG}
\end{document}